# "Coarse" Stability and Bifurcation Analysis Using Stochastic Simulators: Kinetic Monte Carlo Examples.

Alexei G. Makeev[1,#], Dimitrios Maroudas[2] and Ioannis G. Kevrekidis[1,*]

[1]Department of Chemical Engineering, Princeton University, Princeton, NJ 08544

[2]Department of Chemical Engineering, University of California, Santa Barbara, Santa Barbara, CA 93106

[#]Permanent Address: Moscow State University, Faculty of Computational Mathematics and Cybernetics (BMK) Moscow, 119899, Russia

[*]To whom correspondence should be addressed.

## Abstract

We implement a computer-assisted approach that, under appropriate conditions, allows the bifurcation analysis of the "coarse" dynamic behavior of microscopic simulators without requiring the explicit derivation of closed macroscopic equations for this behavior. The approach is inspired by the so-called time-stepper based numerical bifurcation theory. We illustrate the approach through the computation of both stable and unstable coarsely invariant states for Kinetic Monte Carlo models of three simple surface reaction schemes. We quantify the linearized stability of these coarsely invariant states, perform pseudo-arclength continuation, detect coarse limit point and coarse Hopf bifurcations and construct two-parameter bifurcation diagrams.

**Introduction:** Consider a chemically reacting system for which a lumped model, *i.e.* a set of coupled ordinary differential equations (ODEs) based on conservation laws and kinetic closures provides an adequate representation. Two distinct computational approaches to the study of the model's behavior exist. The first is *direct simulation*: prescribing operating conditions (parameter values), initial conditions, and integrating forward in time. The second is *numerical bifurcation theory*: steady states, both stable and unstable, are found through contraction mappings like the Newton-Raphson method; arclength continuation is used to follow the dependence of steady states on parameters [1,2], and test functions are used to detect and accurately compute and continue bifurcation points, *e.g.* turning points, associated with steady state multiplicity and hysteresis, as well as Hopf bifurcations associated with the onset of oscillations. This approach, which uses the same physical model as the direct simulation, has been often proven more computationally efficient in characterizing aspects of the long-term system dynamics. Applied mathematics and numerical analysis are combined to construct computational superstructures (numerical bifurcation algorithms) "around" the dynamic mathematical model, which analyze the model's behavior much better than direct simulation.

If the macroscopic lumped model is not explicitly available in closed form, we often have available microscopic, molecular-level dynamic models capable of describing the physics at different levels of coarse graining. Molecular Dynamics (MD), Monte Carlo (MC) and kinetic theory based (Lattice Gas (LG), Cellular Automata (CA) and Lattice Boltzmann (LB)) schemes - as well as hybrid schemes- have been successful in describing the dynamic behavior of many physical, chemical and engineering systems. This is mostly accomplished, however, in a manner analogous to the *direct simulation* we mentioned above for systems of ODEs; the analogue of numerical bifurcation calculations is essentially lacking. MC models do not generically have strict steady states, and the exact steady states of MD models are quenched states that correspond to minimum energy maps of macroscopic steady states. The latter are steady states of the *moments* of the molecular distribution we evolve, such as concentration, the zeroth moment. We will use the term "coarse steady states" for these. Typically, in order to perform bifurcation calculations we first need to derive closed macroscopic equation(s) for moments of the microscopic distributions, and then analyze these closed equations. The purpose of this paper is to bypass the step of deriving explicit closed macroscopic equations (the time-honored continuum conservation laws). We only assume that such laws *exist, and close at some level of* *truncation* (*e.g.*, in terms of moments of the microscopic distributions); we then proceed to analyze these equations without deriving them explicitly. In this paper, we implement this approach for three chemically reacting systems described microscopically by Kinetic Monte Carlo (KMC) simulation.

The computational entity that makes this task possible is the so-called "coarse time stepper", and the entire procedure is inspired by time-stepper based numerical bifurcation methods for continuum problems [3,4,5]. The idea is to use a time-stepper *i.e.* a subroutine that integrates the system equations, or more generally evolves the system state in time. The time stepper is not used in the traditional "Picard iteration" way, which consists of taking the result of the integration for a finite time interval, use it as an initial condition, and repeat the process, "bootstrapping" the dynamics into the future. Instead, using nearby initial conditions and ideas from iterative linear algebra and separation of time scales, one can carry out what has been called [6] Newton-Picard iteration and consists of four steps:

- identify a (low–dimensional) subspace P in which the dynamics are slow, even possibly unstable;
- approximate the "slow" Jacobian, *i.e.* the corresponding linearization in P;
- use Newton's method in this subspace combined with integration (Picard iteration) in its orthogonal complement Q, and thus
- locate iteratively steady states $\mathbf{x}^*$ as fixed points of the time-$\tau$ map $\mathbf{F}_\tau(\mathbf{x})$, *i.e.* as solutions of $\mathbf{x}^* - \mathbf{F}_\tau(\mathbf{x}^*) = \mathbf{0}$, the result of integrating the system equations with initial condition $\mathbf{x}$ for time $\tau$.

The steady states thus located are not found because they are *stable*, but because they are *dynamically invariant*: it is possible to use a *forward* in time integrator to find, in this way, saddle points or sources too. The construction of continuation and multiparameter continuation algorithms directly follows. In a number of publications we have described the use of this approach to the coarse bifurcation analysis in distributed microscopic systems [7,8,9,10] (see also [11]); there the examples involved mostly Lattice-Boltzmann LB-BGK, kinetic theory based methods. In this paper we will focus on lumped models, stressing the stochastic nature of the time evolution rules. The treatment of distributed stochastic problems using these ideas and the ones presented in [7,8,9] directly follows.

**The models:** To illustrate our approach we will use microscopic KMC simulations of three simple surface reaction models for which we know the mean field (MF) evolution equations for the coverages $\theta_i$ of species (i) on the surface. **Model 1** results from a drastic simplification of the kinetics of NO reduction by $H_2$ on Pt and Rh surfaces [12]:

$$d\theta/dt = \alpha(1 - \theta) - \gamma\theta - k_r(1 - \theta)^2\theta, \qquad (1)$$

where $\alpha = 1$ and $\gamma = 0.01$ are the rate constants for adsorption and desorption of NO, respectively, and $\theta$ represents the coverage of adsorbed NO. The reaction term takes into account the fact that two adjacent adsorption sites are required for adsorption of $H_2$. We take the reaction rate constant $k_r$ as our bifurcation parameter. In a certain range of $k_r$, $3.96 \leq k_r \leq 26$ the model exhibits bistability. This model is chosen because of its simplicity: the bifurcation behavior can be studied analytically. **Model 2** is the standard model for the A+ 1/2 $B_2$ → AB reaction. The model is a simplified kinetic description of the CO oxidation reaction given by the evolution equations

$$d\theta_A/dt = \alpha\theta_* - \gamma\theta_A - 4k_r\theta_A\theta_B, \qquad (2a)$$
$$d\theta_B/dt = 2\beta\theta_*^2 - 4k_r\theta_A\theta_B, \qquad (2b)$$

where $\theta_* = 1 - \theta_A - \theta_B$. The values of $\alpha$, $\gamma$ and $k_r$ are fixed ($\alpha$=1.6, $\gamma$=0.04, $k_r$=1), and $\beta$ is the distinguished bifurcation parameter. This model also exhibits bistability. **Model 3** is chosen because it exhibits supercritical Hopf bifurcations and oscillatory behavior, through introducing to the description of Model 2 an inert site-blocking adsorbate with a reversible adsorption step [13,14]:

$$d\theta_A/dt = \alpha\theta_* - \gamma\theta_A - 4k_r\theta_A\theta_B, \qquad (3a)$$
$$d\theta_B/dt = 2\beta\theta_*^2 - 4k_r\theta_A\theta_B, \qquad (3b)$$
$$d\theta_C/dt = \mu\theta_* - \eta\theta_C, \qquad (3b)$$

where $\theta_* = 1 - \theta_A - \theta_B - \theta_C$. Here we use $\mu$=0.36 and $\eta$=0.016. The values of $\alpha$, $\gamma$ and $k_r$ are fixed, being the same as in model 2. Using $\beta$ again as the bifurcation parameter, one can find two points of supercritical Hopf bifurcations (at $\beta \approx 20.3$ and $\beta \approx 21.2$).

**Kinetic MC Simulations.** We will use KMC simulations to approximate the solution of the corresponding master equation, which describes the evolution of the probability (PDF) of finding the system in a certain configuration. We assume here that the adsorbed layer is randomly distributed due to rapid diffusion of adsorbed species (hopping rates of all adsorbates species are taken to be many orders of magnitude greater than other relevant rates); modifications to the process as surface diffusion becomes slower will be considered in a future publication. We construct a system with N adsorption sites; $N_i(t)$ is the number of species "i" entities. Transition probabilities per unit time for process "j", $R_j$, are dictated by the corresponding terms in the MF models, where $\theta_i = N_i/N$. In our case, the KMC algorithm represents a stochastic implementation of the MF equations [15,16]. The time $\Delta t$ that the system spends in a current configuration is $\Delta t = -\ln(\xi)/R_{tot}$, where $\xi$ is a random number uniformly distributed on (0, 1), and $R_{tot} = \Sigma_j R_j$ is the total transition probability per unit time. Another random number is used to select which reaction (from the set of all possible reactions) will occur with a probability proportional to its rate, and the number of molecules of each species participating in this reaction is changed according to the stoichiometric coefficients. Thus, to simulate each microscopic event, two random numbers have to be generated. If N tends to infinity, the results of the deterministic (MF) and stochastic models coincide. However, the amplitude of coverage fluctuations increases near turning points and Hopf bifurcation points.

**The time-stepper and its uses.** If the continuum conservation equations above are available, they can be inserted in numerical bifurcation algorithms, like the ones implemented in software such as AUTO [17] or CONTENT [18], which then perform the computer-assisted bifurcation analysis. Suppose that for some reason the equations are not available, but we do have a "black box" timestepper $\mathbf{F}_\tau(\mathbf{x}_0)$ for each initial condition $\mathbf{x}_0$. This timestepper might have been constructed by discretizing time through, say, a Runge Kutta or a backward difference procedure, writing a subroutine that, given an initial condition $\mathbf{x}_0 \equiv \mathbf{x}(t=0)$ integrates the equations until the reporting horizon $\tau$ and produces the result $\mathbf{F}_\tau(\mathbf{x}_0) = \mathbf{x}(t=\tau)$. For what we will now do the integrator does not necessarily have to be accurate, nor numerically stable [3,19]. Then it is still possible to perform computer-assisted, but now time-stepper based, numerical bifurcation analysis as follows:

(a) Perform a Newton-Raphson solution of the fixed point equations $\mathbf{x} - \mathbf{F}_\tau(\mathbf{x}) = \mathbf{0}$ with numerical derivatives, i.e. derivatives evaluated by integrating nearby initial conditions for the same time $\tau$ (we have found centered differences to perform satisfactorily).

(b) Analyze stability of the located fixed point through the eigenvalues of the numerically approximated linearization **DF** upon convergence. For a "perfect" integrator the eigenvalues of **DF** are $\nu_i = \exp(\lambda_i \tau)$ where $\tau$ is the integrator time-reporting horizon and $\lambda_i$ are the corresponding eigenvalues of the ODE linearization. Asymptotic stability is guaranteed if all $\nu_i$ are within the unit circle in the complex plane. The eigenvectors of the linearization of the ODEs are retained by the linearization of the "perfect integrator" **DF**. Different integration schemes will have the same steady state, often the same linearization eigenvectors, and eigenvalues from which the stability of the time-stepper fixed point and (separately) the stability of the original ODE steady state can be extracted.

(c) Using numerical derivatives of the time-stepper with respect to parameters, as well as finite difference approximations of second derivatives, it is possible to construct pseudo-arclength continuation codes, turning point and Hopf bifurcation detection codes, as well as (through further numerical differentiation) higher codimension bifurcation detection codes.

It should be noted that if one deals with large systems of ODEs or discretized dissipative PDEs, then the Recursive Projection Method [3] can be used to perform Newton-Raphson iterations only in a low-dimensional, adaptively identified, slow subspace; integration itself can "kill" the remaining strongly (fast) stable directions. We refer the reader to [3,20,21,22] for numerical analysis aspects (e.g. convergence proofs) for these techniques.

The rationale behind performing bifurcation analysis based on microscopic dynamical simulators is that we can replace the timestepper of the coarse equations with the **coarse time-stepper** constructed through the following procedure, which operates through the microscopic (*e.g.* KMC) simulator on the microscopic system description. The procedure consists essentially of the following five steps:

(a) Choose the statistical averaging in describing the coarse problem (in our case, average coverage, $\theta_i$) and a **(coarse) initial condition** for this statistical description (averaged quantity) **q**.

(b) Define a *lifting* operator **m(q)** that constructs *microscopic* initial conditions (distributions) **Q** consistent with the *coarse* initial condition **q** (distributions conditioned on some of their lower moments – in this case the zeroth moment). This operator is **not uniquely defined**; actually, we may need to come up with not just one but possibly a reasonable ensemble of microscopic initializations of **Q** consistent with **q**. Every time this is initialization step is performed, information is "injected" into the problem: in addition to the moments of the distribution that we do initialize consistently, we also prescribe some, in principle arbitrary, initial conditions for all other moments of the distribution.

(c) Evolve the initial microscopic distribution through the microscopic evolution rule for time $\tau$; this time –here, the reporting horizon of the KMC simulation—corresponds to the time-reporting horizon of the coarse integrator above. For reasons of variance reduction, we will need to perform *several* Monte Carlo integrations of the same initial condition(s) constructed above.

(d) Average the results among Monte Carlo evolutions of the consistent initializations from the ensemble constructed above; the averaging is done over realizations, over initial conditions, and conceivably over microscopic, "fine" time. For distributed problems, the averaging should also be performed over microscopic, "fine" space [7,8]. We then "project" (we use the term *restrict* and the symbol **M**) the results to the coarse description (moments **q**$(t=\tau)$ = **M(Q**$(t=\tau)$**)** of the microscopic description – the distributions **Q**).

We expect that this "lift – evolve – average - restrict" procedure gives us a good estimate of the time stepper for the closed macroscopic description, even though we do not have this description explicitly available. This is because we expect that –for the type of problems we are interested in— the higher order moments of the microscopic description evolve very fast to functionals of the low, "governing" moments that we initialized consistently. Alternatively, we can say that the higher order moments become *slaved to* the lower order ones, and that, therefore, the problem is a singularly perturbed one in the "moment hierarchy" description. This concept resonates with the theory of inertial manifolds and approximate inertial manifolds for dissipative PDEs [23,24,25]. The associated separation of time scales implies that, if the time horizon of the microscopic evolution is not too short (so that errors in the initialization of unknown moments get "healed") and also not too long (so that the initial conditions are "forgotten", and/or the various realizations diffuse irreparably over the overall PDF, losing phase information), the procedure is indeed a good approximation of the unavailable closed evolution equation for the low, governing moments. This is then an "unavailable model motivated" co-processing of the Monte Carlo simulation.

> (e) We now simply substitute the coarse time-stepper in the above described, continuum time-stepper based, numerical bifurcation procedure. The computational superstructure "does not realize" that the time-stepping results come from appropriately initialized and averaged microscopic simulations, and not from a continuum equation. Therefore, it proceeds to perform the bifurcation analysis of the (unavailable in explicit form) closed continuum equation.

We will now demonstrate that this substitution works for our examples. We use the Monte Carlo simulation subroutine (via a computational superstructure that constructs and exploits the coarse time-stepper) to compute and continue stable and unstable *coarse steady states* (*i.e.* coarsely invariant, stationary solutions of the moments of the distribution). We then perform coarse stability analysis, detect and continue coarse bifurcations – in short, perform the analysis of the, in principle, unavailable coarse (macroscopic) equation. The fact that, in this paper, we happen to know what this equation is, allows us to check the coarse Monte Carlo results. We will do this using system sizes that support quite large fluctuations, away from the practically deterministic limit where the number of particles goes to infinity.

**Results:** Figure 1a shows the coarse-time-stepper-based computation of the one-parameter bifurcation diagram of the KMC simulation of Model 1 with respect to the reaction transition probability per unit time, $k_r$. The number of available sites (system size) was $N=100^2$, and $N_{run}$ =2000 realizations were performed for each coarse initial condition for variance reduction purposes. This variance reduction allowed us for enough precision in the computation of numerical derivatives of our coarse time stepper, for the Newton-Raphson fixed point algorithm to (in an almost always sense) converge. Typically we used perturbations of the order of $10^{-2}$ in our centered difference computation of numerical derivatives, and declared convergence of the Newton-Raphson when the residuals became $O(10^{-5})$. The time reporting horizon of the Monte Carlo simulations correspond here to $\tau = 0.025$, a time long enough compared to individual Monte Carlo events (a reporting horizon involved typically more than 1000 Monte Carlo steps) and short enough compared to the macroscopic time scales of the problem. These latter time scales can be estimated from the eigenvalues of the linearization of the mean field equations around their steady states.

It is clearly possible to compute both stable and unstable coarse steady states. In Fig.1(a) the solid line is the result of continuation of the MF equation, while the points are the results of converged Newton-Raphson iteration for the coarse time stepper. Upon convergence of the Newton-Raphson algorithm (which is limited by the variance reduction in our coarse time-stepping) we have obtained the coarse Jacobian as a byproduct of the process. For this one-dimensional system, this coarse Jacobian is just one derivative: the derivative of the right hand side of the equation for the average coverage with respect to this average coverage evaluated at the steady state. This quantifies *coarse local stability* of the coarse steady state. The coarse eigenvalue is compared with the analytically computed one in Fig.1(b). Furthermore, through the implementation of pseudo-arclength continuation based on the coarse time stepper, it is possible to compute complete one-parameter bifurcation diagrams [Fig.1(a)] and even two-parameter continuations of turning points of the one-parameter diagrams [Fig.1(c)].

It is interesting to observe the slightly inaccurately detected upper turning point in Fig.1(a). To some extent, the weights in the pseudo-arclength continuation affect this computation. More than anything else, however, this is a result of insufficient variance reduction; as more copies and larger system sizes are integrated over a short time, the accuracy in the detection of the turning point increases. This brings up, however, a point that we consider a truly important feature of this approach. It is well known that the fluctuations in a dynamic simulation are going to become more and more pronounced in the neighborhood of critical points, such as turning points, or Hopf points. This will happen if we make long simulations for fixed parameter values very close to these transitions. This is "echoed" in some sense in the fact that the continuum simulation (which really breaks down at these values, in the sense that it does not provide a good description of the microscopic behavior any more) slows down immensely at near-critical parameter values. This is not observed, however, in our simulations; the reason is that we perform finite time simulations, and we actually do not perform them at fixed parameter values. Indeed, we solve an augmented "coarse" system, constructed through the arclength continuation. This way, we update an "augmented state" that consists of *both* the coarse steady state *and* the parameter value. In the present example, this is done in order to find the point $(\theta, k_r)$ at a given arclength distance $(\theta-\theta_0)^2 + (k_r - k_{r,0})^2$ from the previous one $(\theta_0, k_{r,0})$ on the bifurcation curve. What is a turning point for the "constant parameter" simulation (marginally stable for the MF description) becomes a *regular* fixed point for the augmented system in state×parameter space – so critical fluctuations do not have time to develop – the parameter changes along with the coarse state and prevents their buildup. We can think of this new procedure (performing arclength continuations for the coarse time stepper) as a new "coarse motivated" augmented ensemble for the Monte Carlo simulations. Inspired by path-following methods we assign to this procedure the term "pathostat".

The reason for the much crisper curve definition in Fig.1(c) is that much more computational effort was given to variance reduction: the system size was $N=(1000)^2$ and the number of realizations averaged was $N_{run}=1000$ also. A larger system size is dictated by variance reduction considerations: second partial derivatives had to be numerically approximated here, since we were continuing a codimension-one coarse turning point curve past codimension-two coarse bifurcation points: here "turning points of turning points" *i.e.* cusp points. It is again interesting to realize that Fig.1(c) embodies an algorithm that **converges** (almost always, almost quadratically) in (coarse) state×parameter space to what, for the non-augmented system, is a transition point (a marginally stable point for the continuum model). The two-parameter continuation algorithm involves yet another "coarse motivated" ensemble for doing KMC simulations: one that tracks transition points (folds) in two-parameter space; one might call it a "foldostat; "coarse motivated" ensembles for the location and multiparameter continuation of other transition points (like the Hopf bifurcation of Model 3) easily follow from the combination of the coarse time stepper and traditional bifurcation algorithms . It may be a good idea to devise a consistent nomenclature for these coarse bifurcation-inspired KMC co-processing algorithms thinking of them as alternative simulation ensembles. The "invaded cluster" algorithm of Machta and coworkers [26] is the closest to a "fold-detection alternative ensemble" that we have found in the current literature.

Figure 2 shows the same type of coarse-timestepper based bifurcation diagram for model 2; the projection of the bifurcation diagram on both coarse variables, $\theta_A$ and $\theta_B$ is shown in Fig.2(a). Again, there is very good agreement, as expected, with the mean field results. The slowest of the two eigenvalues, $\lambda_1$, of the coarse linearization is shown in Fig.2(b). Notice that, as in Fig.1, we can indeed converge on *coarsely marginally stable* turning points, where the coarse eigenvalue of the linearization is zero.

Figure 3 shows the results of numerical bifurcation analysis based on the coarse time stepper of Model 3. Figure 3(a) provides an indication of how difficult it is to pinpoint Hopf points, *i.e.*the onset of coarse time-periodic states, with some accuracy from direct simulation: it shows the time series (just before the first Hopf bifurcation of the mean field model), of two single Monte Carlo realizations, one for $N=200^2$ and one for $N=1000^2$. Notice how both of them have rather irregular, finite amplitude oscillations. Figure 3(b) shows the corresponding one-parameter bifurcation diagram: the diagram exhibits two coarse Hopf bifurcations and indicates that our algorithm succeeds in finding the coarse unstable states between the two coarse Hopf points. The three eigenvalues of the coarse linearization of the three-equation Model 3 in this case are shown in Fig. 4. Clearly, the coarse-timestepper based bifurcation calculations do a much cleaner and definitive job in predicting the coarse bifurcation than simple simulation would. Notice how both the real and imaginary parts of the coarse eigenvalues are captured through the procedure. It is quite straightforward (in the same way that we created a co-processing scheme for the location and continuation of turning points, above) to construct now a superstructure (coarse bifurcation code, alternative ensemble) to locate and continue coarse Hopf bifurcations in multiparameter space. Once more, the convergence of all these algorithms is in an "almost always" sense. Furthermore, we make sure that the timestepper will never run for so long a reporting horizon at a critical parameter value for the coarse description to break down.

Figure 5(a) illustrates an important characteristic of the coarse time-stepper that should be discussed. Several different simulations of the system are plotted in the regime between the two Hopf bifurcations, where the prevailing mean field solution is a stable limit cycle. The black solid curve is an accurate simulation of the mean field continuum equations. The cyan, broken curve shows **a single** KMC realization of the dynamics: the amplitude and frequency are there, but only on average and in a noisy manner. The blue solid curve is the result of running $N_{run}$=1000 KMC simulations from the same microscopic initial condition: they track, as one might expect, the bold line, for a longer time than a single noisy realization, but the curve loses its quality quite fast: it will eventually more or less plateaus. The reason for this is the fact that one loses phase information fast, and what we see is a sampled average over the ultimate Probability Density Function (PDF). The shape of this PDF plotted as a function of the first two moments, $\theta_A$ and $\theta_B$, will correspond to a "volcano"-like picture: the probability is larger to find the system along the "rim" of the crater, that represents the projection of the limit cycle.

The important element is a fourth (yellow solid) curve, obtained in the following way:

(a) Start with a coarse initial condition ($\theta_A$, $\theta_B$, $\theta_C$).
(b) Lift this initial condition to a consistent microscopic one ($\Theta_A$, $\Theta_B$, $\Theta_C$).
(c) Run the same number (1000) copies of the Monte Carlo evolution **but for a reasonably short time**. This time should be long enough for the possible errors in initializing higher moments to "heal"; but short enough so that phase information is not irreparably lost, and memory of the initial condition is retained. Clearly, there has to exist a range of such reporting horizons for which the procedure may be successful, and all of them should give the same coarse fixed points, the same coarse limit cycles, and, more generally, the same coarse global attractor. The idea is that this time stepper can conceivably be derived from a semi-group **on** the space of moments, *i.e.* that a generator for this conceptual coarse semigroup exists.
(d) Average and restrict the microscopic results to governing moments (**q**).
(e) Go to (b)

This "lift – evolve – average - restrict" procedure helps the Monte Carlo simulation track the effective coarse dynamics very successfully, and for a long time. The important quantity here is the so-called "density of collapses" (or frequency of restrictions) in time. As we explained above, if the time reporting horizon of the time-stepper is too short, the errors in lifting do not get time to "heal". If on the other hand the reporting horizon is too long, then (as in Fig. 5) we lose phase information, and the ability to approximate the effective continuum model. Fig.5(b) shows the same information in a phase space representation, *i.e.* projected onto the zeroth moments $\theta_A$ and $\theta_B$. The black solid line shows the same mean field trajectory. The cyan points are the result of 1000 Monte Carlo integrations reported every 2 time units (seconds) up to time =2000 (a million points, representative of the ultimate PDF for a $N=200^2$ system). The yellow curve is the result of our "lift-evolve-average-project" approach for 1000 KMC copies. A coarse initial condition is lifted to a consistent microscopic one, 1000 KMC copies are performed for 5 seconds, and then are averaged and restricted, and the procedure repeats up to time=5000 (1000 iterations). The yellow points shown (starting from the top left and going down and to the right) are the results of this procedure at a number of times instances (see caption) . The red "clouds" around the points show the scatter of the 1000 Monte Carlo simulations since the last lifting. Due to the frequent "collapses" in our procedure, phase information on the limit cycle is not lost; a limit cycle is indeed obtained with a well-defined period. Notice also the dark blue points: they are the result of 1000 MC simulations that started at the same initial condition as our procedure, but run for time 5000 without liftings and restrictions: clearly, they have diffused over the entire PDF, phase information has been lost, and all one sees for their average is a practically fixed point (the triangle at the middle of the graph). This triangle does *not* represent the unstable steady state in the middle of the limit cycle (which we have found through the coarse Newton-Raphson timestepper approach above). It represents the average of the entire stationary distribution; it can be thought of as a weighted mean over the limit cycle for deterministic systems, and actually is a weighted (with the measure) mean over the entire PDF for the stochastic one. The limit cycle obtained through our procedure (as a set in moments space) and its period are close to the limit cycle and the period of the mean field problem. The slight difference depends on the details of the "lift-evolve-average-restrict" procedure as well as the actual size ($N=200^2$) of the problem. We are currently using this version of the coarse time-stepper to compute the coarse limit cycle, its coarse period and its coarse Floquet multipliers as a fixed point of a Poincaré map in moments space.

It is important to note here that there appear to be two senses of "infinite time" in the microscopic evolution of distributions: (a) the correct "infinite time" one, which misses phase information and "diffuses" the distribution over configuration space and phase space – we just see the ultimate PDF; and (b) the "macroscopic equation" infinite time, where one tries to find attractors and instabilities. The point of this discussion is that the objects we are discussing here (the ω-limit set of the macroscopic equations and its dependence on parameters) may be only a "summary" of the PDF; but it is a very practically useful summary, for whose analysis incidentally a lot of wonderful applied mathematical/scientific computing techniques exist. What we have proposed in this paper is a computer-assisted bridge that makes the analysis tools for the macroscopic equation directly useful to the microscopic description, without the intermediate bottleneck of the derivation of the macroscopic equation.

**Discussion:** There are, of course, several caveats in the procedure we have outlined above; we have not described under what conditions it is expected to work. We believe, however, that under the assumptions used to derive coarse closed macroscopic equations it may be possible to show that this approach does indeed provide (in an "almost always" sense, as in stochastic approximation algorithms [27,28]) a good approximation of the bifurcation analysis of the unavailable closed coarse equations. One of the interesting points of this procedure, from the computational point of view, is that it dictates the computation of several, uncoupled microscopic simulations of "the same" coarse initial conditions; as such it is ideally suited for massively parallel computation.

It is possible to extend this methodology to simulations of spatially distributed systems; examples of the coarse analysis of non-stochastic, kinetic theory based schemes (LB-BGK models of reaction diffusion problems and of multiphase flows) have appeared [7,8], while work on chemotaxis [29] and lattice gas problems is in progress. It is also straightforward to extend the coarse steady state location algorithms to compute coarse limit cycles : one uses a Newton-Picard approach to find fixed points of the coarse Poincaré map; the important point is that integration over the long times required for a Poincaré return must practically be done through the lift-evolve-restrict approach advocated above. The procedure corresponds to a shooting solution of a boundary value problem in time, and will have no problem converging to unstable coarse limit cycles.

It is important to notice that our coarse time stepper is not only useful in locating the infinite-time behavior, such as steady states and bifurcation diagrams, but can also be used to perform what we call "coarse integration" through the so-called Projective Integration techniques [11,8]. We have also used this approach to perform the bifurcation analysis of "effective medium" behavior for complex media without first explicitly obtaining the effective equations [10]; for the "coarse" analysis of chains/lattices of oscillators (in particular, models of coupled neurons), again without the explicit derivation of an effective medium equation; and for the coarse analysis of biased random walks as well as stochastic models of cell motility and chemotaxis [29].

How does our procedure fail ? Although this is the subject of detailed study in a forthcoming publication, it is worth including some discussion here, since it is one of its more important characteristics. In thinking about the macroscopic, moments-level description of the evolution of a distribution, we can conceive of three regimes: Regime A, where a particular low-moments macroscopic equation exists, closes, and is an adequately accurate representation of reality; this means that not only the level at which the description closes has been successfully chosen, but also the particular right hand side we have arrived at based on assumptions/approximations does a good job in representing the coarse dynamics of the distribution. In this regime, the "restriction" of the true dynamics to moments, the particular macroscopic equation, and our coarse approach, all give the same results. This is followed by regime B, where *a* macroscopic equation exists and closes at the same moments level; but the right hand side we can derive is not so accurate any more. In this regime the macroscopic equation predictions will be less good approximations; but the results of our approach using the coarse timestepper, and conditioning initialization at the same coarse level (number of moments), will –we believe- still give the same results with the true system (the restriction of microscopic simulations to moments). Finally, there is regime C, where the macroscopic equation is totally inaccurate, and where the system does **not** close at the same level of description any more – more moments are necessary to obtain a deterministic coarse model. These additional moments, which could be thought of as a hierarchy of singularly perturbed variables, with very fast dynamics up to now, have become slow enough so that they must be included in the simulation as independent variables. In this case the "previous" macroscopic equation gives uncontrolled, wrong results. Our procedure, initialized at the "previous" coarse level, will fail to converge – this will appear as the increase in the dimension of the iteratively identified coarse slow subspace, and as the inability to perform sufficient variance reduction to compute precise derivatives no matter what the size of the ensemble used. We are currently investigating the "rate" at which this loss of numerical derivative precision arises, as a "numerical flag" for the necessity to augment the coarse description including new moments. Finally, it is also important to notice that our procedure (exactly the same algorithm, and with the same timestepper) initialized at the *new* coarse description level, *will* work. This is, we believe, a potentially important attribute of this computer-assisted procedure: it forewarns of the failure of a certain description level, suggesting the need for an augmented description; and conveniently fails when it should. This should be contrasted to the closed form macroscopic equations, which do not fail if used beyond their region of validity, but simply produce erroneous results.

Under appropriate conditions, we believe that the approach discussed here can aid the computer-assisted study of a wide class of microscopic deterministic and stochastic models (from MD and hybrid transport problems to MC simulations of kinetics, signaling and motility of cells and pure stochastic PDEs). Although "textbook" examples were studied in this paper (the KMC was built around the MF description), we are currently studying two more realistic examples, involving lateral interactions and diffusion on a lattice based on the same coarse-time-stepper-based approach; our initial results are very promising. We should also note that there are relations as well as differences between our work and work in which approximations of Sinai-Ruelle-Bowen measures of a dynamical system are computed (e.g. [30]).

We do not claim that this approach is applicable in every situation: the assumptions of separation of time scales and slaving are crucial, and, in general, they cannot be easily checked *a priori*. Furthermore, consistent lifting procedures, choices of reporting horizons and representative ensemble constructions are far from trivial. But the approach can be at least tried as a software wrap around existing microscopic or hybrid timesteppers from many branches of science and engineering; we believe that in the appropriate limits, it should be possible to prove that it must work. In particular, the ability to perform RPM-like implementations, where one does not have to compute all of the coarse derivatives by finite differences, should allow the application of the approach to problems that are macroscopically distributed in space.

**Acknowledgements**: The authors would like to acknowledge partial supported through the AFOSR (Dynamics and Control, Drs. Jacobs and King), the National Science Foundation and a Humboldt Foundation prize (I.G.K) as well as the hospitality of the MIT Chemical Engineering Department (DM, sabbatical leave). Discussions with Profs. A. Panagiotopoulos, C. W. Gear, M. Katsoulakis and S. Shvartsman are gratefully acknowledged.

**LITERATURE CITED**

1. H. B. Keller (1997) "Numerical Solution of Bifurcation and Nonlinear Eigenvalue Problems" in <u>Applications of Bifurcation Theory</u>, ed. P. H. Rabinowitz, Academic Press, pp.359-384.
2. E. J. Doedel, H. B. Keller and J.-P. Kernevez (1991) "Numerical Analysis and Control of Bifurcation Problems, (I) Bifurcation in Finite Dimensions", *Int. J. Bifurcations and Chaos*, **1**(3) pp.493-520.
3. G. M. Shroff and H. B. Keller (1993) "Stabilization of Unstable Procedures – the Recursive Projection Method", *SIAM J. Num. Anal.* **30** pp.1099-1120
4. L. S. Tuckerman and D. Barkley (1999) "Bifurcation Analysis for Timesteppers" in *IMA Volumes in Mathematics and its Applications*, E. J. Doedel and L. S. Tuckerman, eds., **119** pp.453-466, Springer, New York.
5. H. Jarausch and W. Mackens, (1987) "Numerical Treatment of Bifurcation Problems by Adaptive Condensation", in <u>Numerical Methods for Bifurcation Problems</u>, W. Kuepper and T. Mittelmann eds., Birkhaeuser, Boston.
6. K. Lust, D. Roose, A. Spence and A. R. Champneys (1998) "An Adaptive Newton-Picard Algorithm with Subspace Iteration for Computing Periodic Solutions", *SIAM J. Sci. Comp.* **19** pp.1188-1209.
7. C. Theodoropoulos, Y.-H. Qian and I. G. Kevrekidis (2000) "Coarse Stability and Bifurcation Analysis Using Time-Steppers: a Reaction-Diffusion Example", *PNAS* **97** pp.9840-9843.
8. C. W. Gear, I. G. Kevrekidis and C. Theodoropoulos (2001) "Coarse Integration/Bifurcation Analysis via Microscopic Simulators: micro-Galerkin Methods", submitted to *Comp. Chem. Eng.* Also can be found as http://www.neci.nj.nec.com/homepages/cwg/UCLA90.pdf
9. C. Theodoropoulos, K. Sankaranarayanan, S. Sundaresan and I. G. Kevrekidis (2001) "Coarse Bifurcation Studies of Bubble Flow Microscopic Simulators", in *Proceedings of the 3d Panhellenic Conference in Chemical Engineering,* pp.221-224
10. O. Runborg, C. Theodoropoulos and I. G. Kevrekidis (2001) "Effective Bifurcation Analysis: a Time Stepper Based Approach", submitted to *Nonlinearity*. Also can be found at http://www.math.princeton.edu/~orunborg/Publications/
11. C. W. Gear and I. G. Kevrekidis (2001) "Projective Integrators for Stiff Differential Equations: Problems with Gaps in their Eigenvalue Spectrum", submitted to *SIAM. J. Sci. Comp.* Also can be found as http://www.neci.nj.nec.com/homepages/cwg/projective.pdf .
12. S.Y. Shvartsman, (1999) Ph.D. Thesis, Princeton University
13. G. Eigenberger (1978) "Kinetic Instabilities in Heterogeneously Catalyzed Reactions 2. Oscillatory Instabilities with Langmuir Type Kinetics" *Chem. Eng. Sci.* **33**, pp.1263-1268
14. R.D. Vigil and F.T.Willmore (1996) "Oscillatory Dynamics in a Heterogeneous Surface Reaction: Breakdown of the Mean-Field Approximation" *Phys. Rev. E* **54** pp.1225-1231

15. D. T. Gillespie (1976) "A General Method for Numerically Simulating the Stochastic Time Evolution of Coupled Chemical Reactions", *J. Comp. Phys.* **22** pp.403-434 (1976)
16. D. T. Gillespie (1977) "Exact Stochastic Simulation of Coupled Chemical Reactions" *J. Phys. Chem.* **81** pp.2340-2361
17. E. J. Doedel (1981) "AUTO: A Program for the Automatic Bifurcation Analysis of Autonomous Systems", *Cong. Num.* **30** pp.265-284 (Proc. 10$^{th}$ Manitoba Conf. Num. Math. And Comp., Univ. Manitoba, Winnipeg, CA); also see versions of AUTO86, AUTO97 and AUTO2000 (see http://indy.cs.concordia.ca/auto/bib/)
18. Yu. A. Kuznetsov and V. V. Levitin (1995-1997) "CONTENT: A Multiplatform Environment for Analyzing Dynamical Systems", Dynamical Systems Laboratory, CWI, Amsterdam (ftp.cwi.nl/pub/CONTENT)
19. E. Koronaki, A. Boudouvis and I. G. Kevrekidis (2001) "Enabling Stability Analysis of Tubular Reactor Models Using PDE/PDAE Solvers", submitted to *Comp. Chem. Eng.*
20. K. Lust (1997) Ph.D. Thesis, Katholieke Universiteit Leuven, Leuven, Belgium.
21. H. von Sosen (1994) Ph.D. Thesis, California Institute of Technology, Pasadena, CA.
22. P. Love (1998) Ph.D. Thesis, California Institute of Technology, Pasadena, CA.
23. P. Constantin, C. Foias, B. Nicolaenko and R. Temam (1988) <u>Integral Manifolds and Inertial Manifolds for Dissipative Partial Differential Equations</u> Springer Verlag, NY.
24. R. Temam (1988) <u>Infinite Dimensional Dynamical Systems in Mechanics and Physics</u>, Springer Verlag, NY.
25. C. Foias, M. S. Jolly, I. G. Kevrekidis, G. R. Sell and E. S. Titi (1988) "On the Computation of Approximate Inertial Manifolds" Physics Letters A 131 (1988) pp.433-436.
26. J. Machta, Y. S. Choi, A. Lucke, T. Schweizer and L. V. Chayes (1995) "Invaded Cluster Algorithm for Equilibrium Critical Points*, Phys. Rev. Lett*. **75**, pp.2792-2795.
27. J. C. Spall (1998) "Stochastic Optimization, Stochastic Approximation and Simulated Annealing" in <u>Wiley Enycopedia of Electrical and Electronics Engineering,</u> J. G. Webester ed. **20** pp.529-542, John Wiley, NY.
28. J. C. Spall (2000) "Adaptive Stochastic Approximation by the Simultaneous Perturbation Method" *IEEE Trans. Automatic Control* **45** pp.1839-1853
29. C. W. Gear, I. G. Kevrekidis, H. Othmer and S. Setayeshgar (2001) *in preparation*
30. M. Dellnitz and O. Junge (1999) "On the Approximation of Complicated Dynamical Behavior" *SIAM J. Num. Anal.* **36**pp.491-515.

**FIGURE CAPTIONS**

Fig.1.

Results of stochastic simulations in combination with a continuation technique for model 1 using $N=(100)^2$, $N_{run}=2000$. Lines correspond to the MF model for stable (solid line) and unstable (dashed) stationary solutions. Surface coverage (a) and the eigenvalue (b) are shown. Circles and squares show surface coverage and eigenvalue calculated by means of the stochastic algorithm. The inset gives typical time-series for different values of $k_r$ demonstrating the increase in the coverage fluctuations while $k_r$ approaches the turning point (at $k_r \approx 3.96$) for $N=(100)^2$ and a single simulation run. The frame (c) shows the two-parameter bifurcation diagram of the saddle-node points, $sn_1$ and $sn_2$, which merge in a co-dimension 2 cusp bifurcation point. Lines (crosses) correspond to the deterministic model (stochastic model with $N=1000^2$, $N_{run}=1000$).

Fig.2.

Bifurcation diagram for model 2 with respect to the parameter $\beta$. Lines correspond to the MF model for stable (solid lines) and unstable (dashed) stationary solutions. Surface coverages (a) and one eigenvalue (b) are shown. The eigenvalue changes its sign at the saddle-node points. Circles and squares correspond to the results of the stochastic simulations using $N=(300)^2$, $N_{run}=1000$.

Fig.3.

Model 3. (a) Time-series for $\beta=20$ near –but before- the Hopf bifurcation point $h_l$, which takes place at $\beta=\beta_1 \approx 20.3$. Initial conditions: zero coverages. Here, the steady state of the MF model (solid thick line) is a stable focus, but the stochastic model for $N=(1000)^2$ (dotted line) and $N=(200)^2$ (solid thin line) appear much more complicated. (b) Bifurcation diagram. Lines correspond to the deterministic MF model. Diamonds mark two points, $h_l$ and $h_2$, of (supercritical) Hopf bifurcation. Circles, squares and triangles correspond to the coarse stationary states of stochastic model with $N=(200)^2$, $N_{run}=2000$. These are obtained as fixed points of the coarse timestepper through our procedure, and have the correct "coarse stability" (they are coarsely unstable between the two Hopf bifurcation points).

Fig.4.

Eigenvalues of the coarse Jacobian matrix (the linearization) at the fixed points of the coarse timestepper of model 3. Frame (a) gives the first eigenvalue, (b) gives the real part of the second and the third eigenvalues, while the imaginary part is shown in (c). Lines correspond to the deterministic model. Squares correspond to the stochastic model with $N=(200)^2$ and $N_{run}=2000$, while triangles in (c) were obtained using $N=(1000)^2$, $N_{run}=1000$.

Fig. 5

Discussion of the "lift - evolve - average - restrict" procedure. Panel (a) shows (1) the deterministic MF simulation of model 3 for $\beta=20.8$ and for zero coverage initial conditions (thin black curve). Superposed to the MF simulation we find (2) a cyan dotted curve (a typical single Monte Carlo realization, $N=200^2$); (3) the instantaneous average of 1000 parallel Monte Carlo realizations that started from the same initial condition (deep blue, notice the loss of phase information); (4) a yellow curve that is the result of our "lift – evolve (1000 MC realizations, for time $\tau=0.5$) – average –restrict" procedure.

Panel (b) shows part of the same results in a (moments phase space) projection. Thin black curve: MF; light cyan points: many MC simulations over a long time – an indication of the ultimate PDF for our $200^2$ system; yellow curve: results of the lift-evolve-average-restrict approach; thick yellow circles: instances of the approach at various indicated times. The red "haloes" show the scatter of 1000 MC simulations between the "lift" and the "restrict" phases of the last step corresponding to the fat yellow circle. The blue triangle at the center of the graph is the average of 1000 MC simulations that started at the same initial conditions as all procedures, and were left to evolve independently up to t=5000. They are clearly scattered over the entire PDF, and all phase information is lost, contrary to the our "yellow curve" procedure.

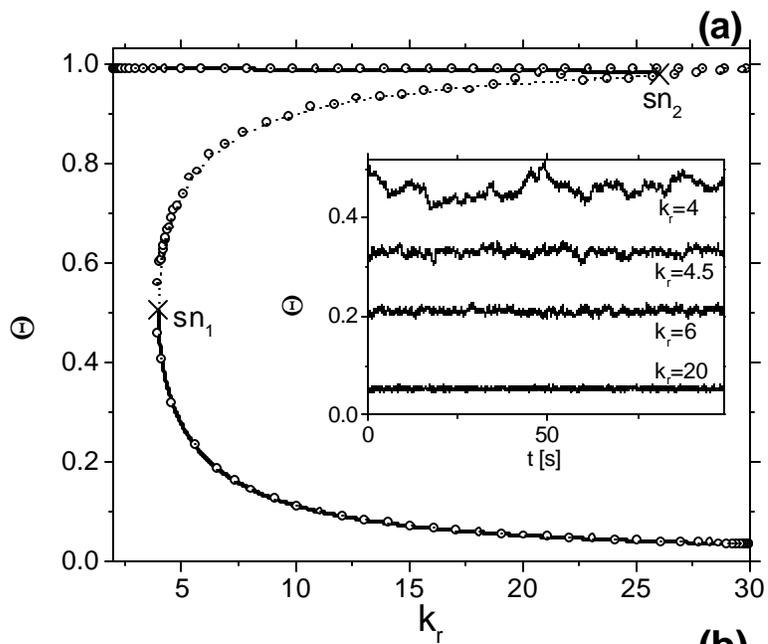
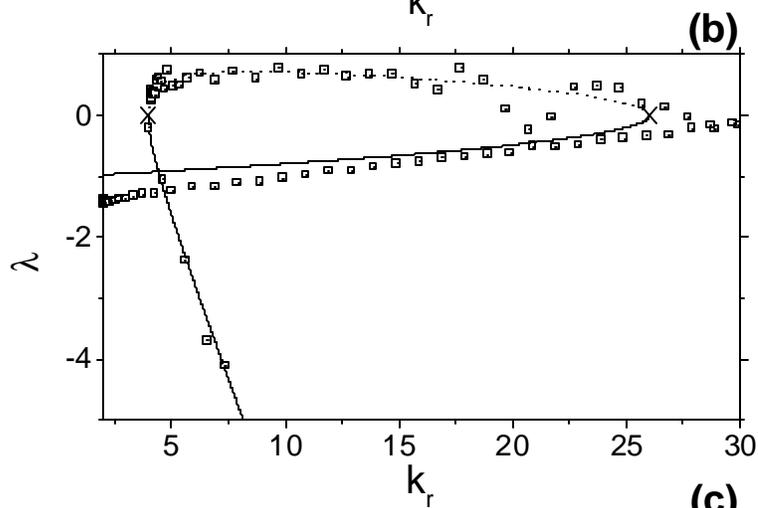
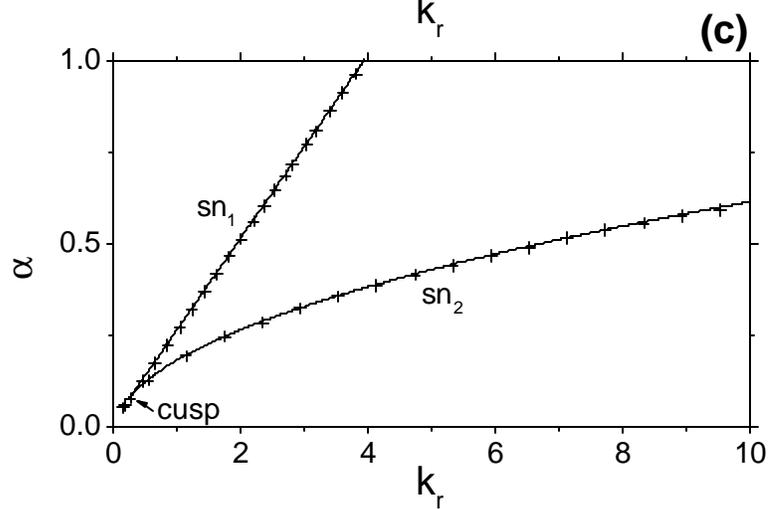

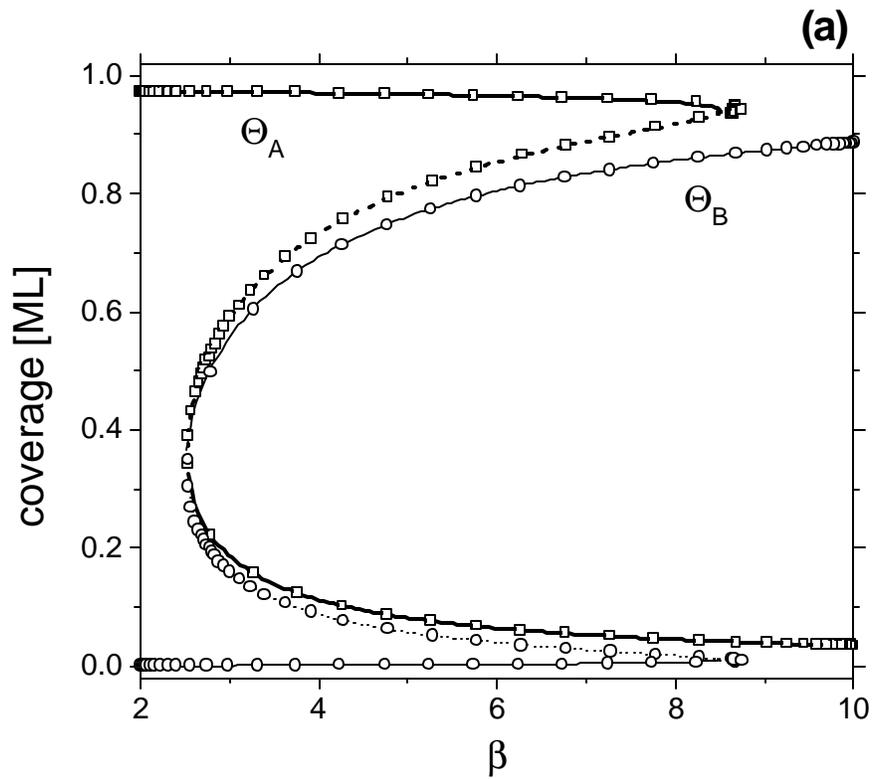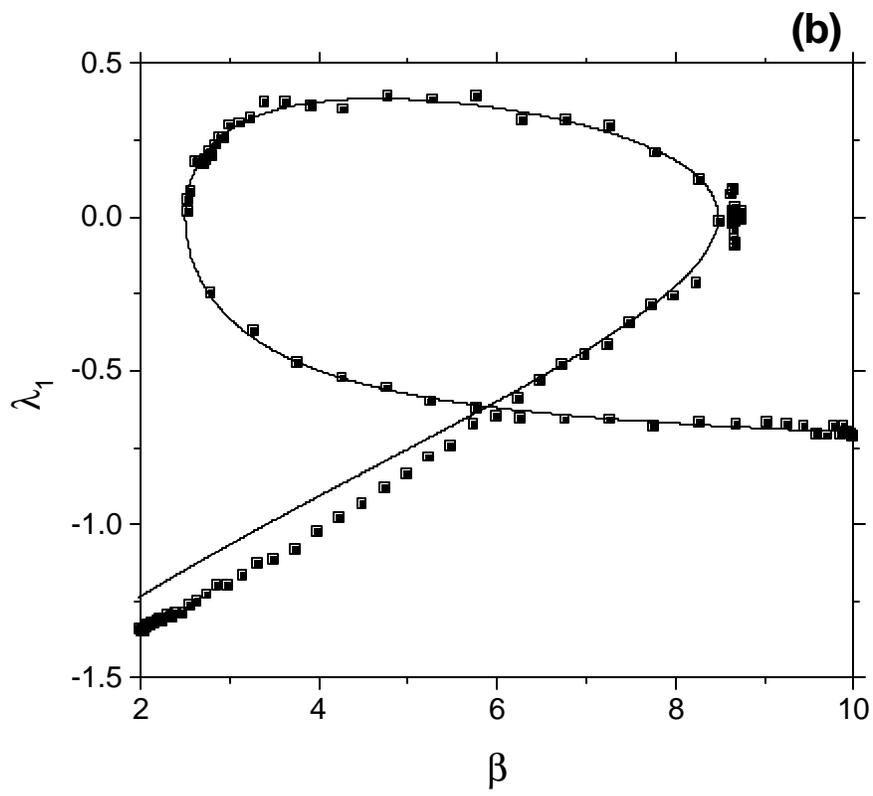

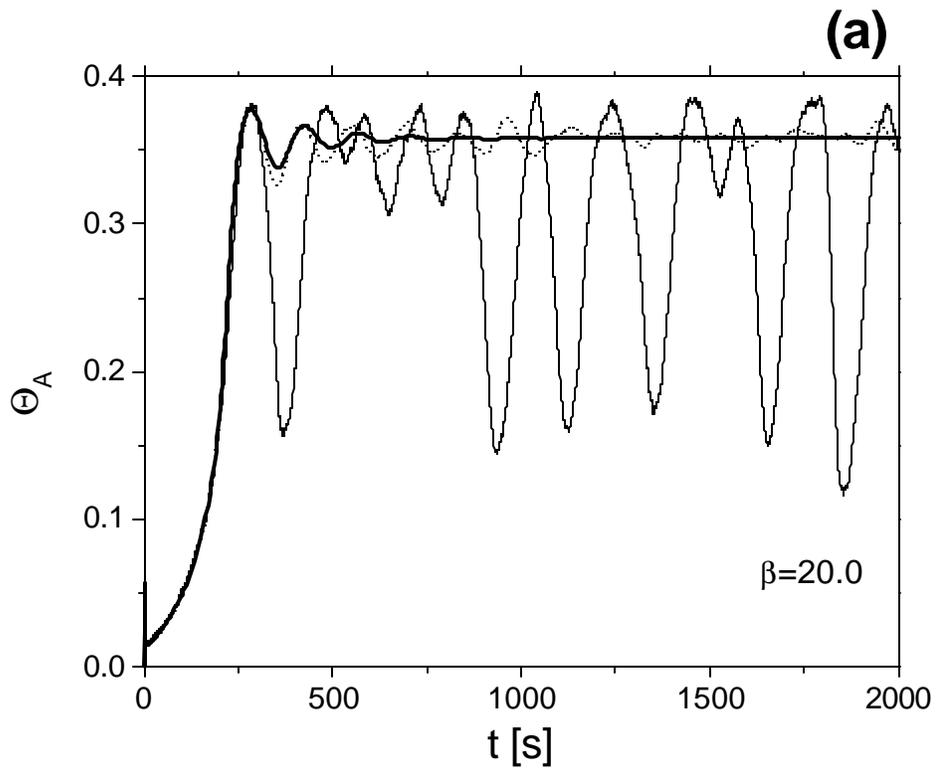

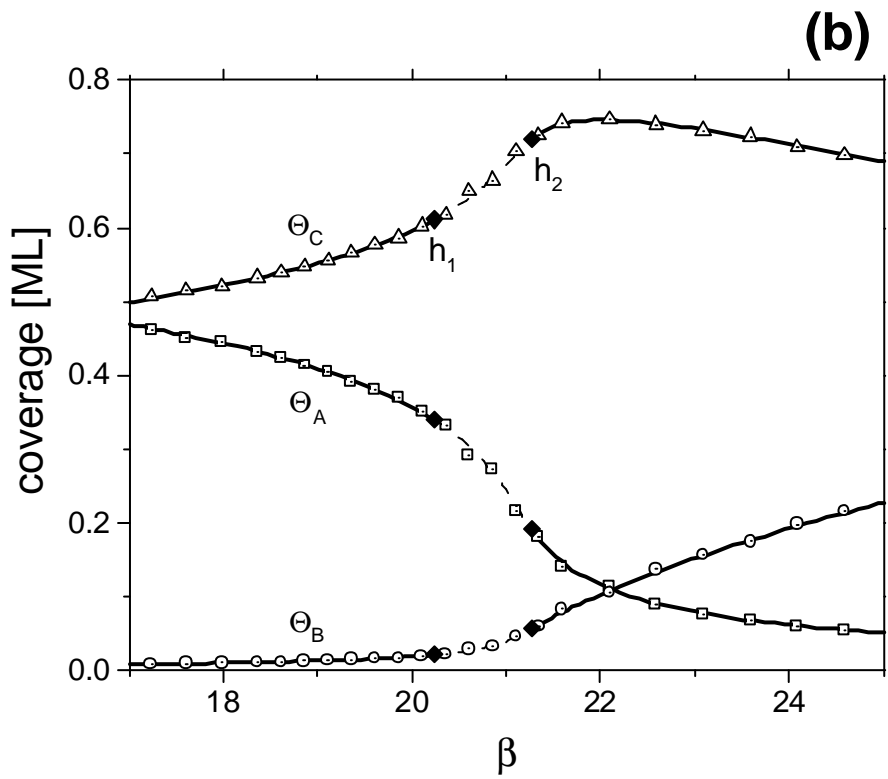

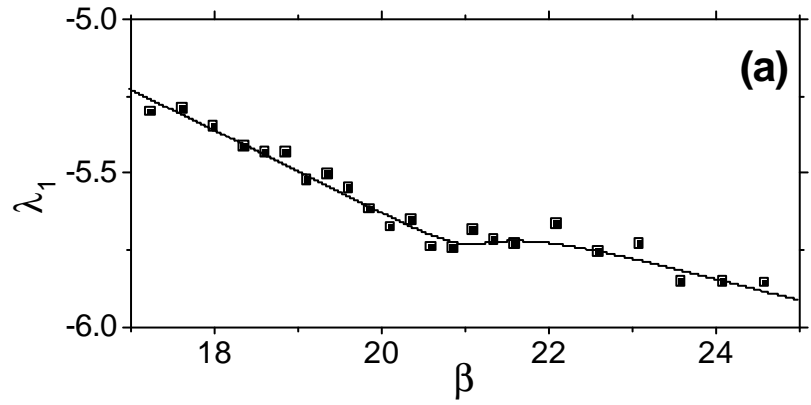

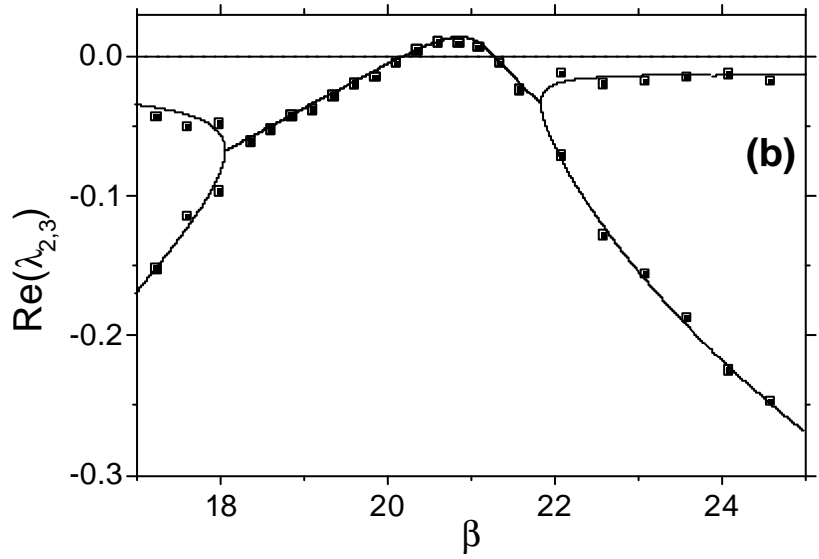

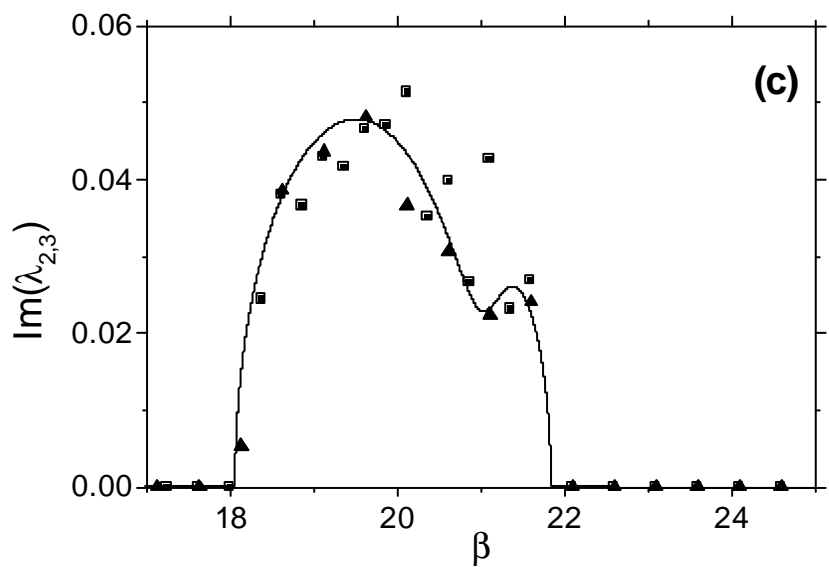

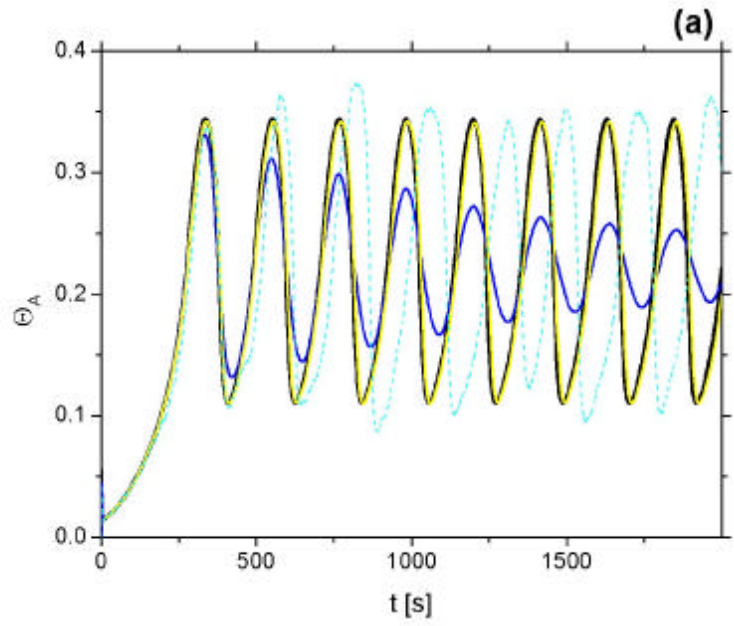
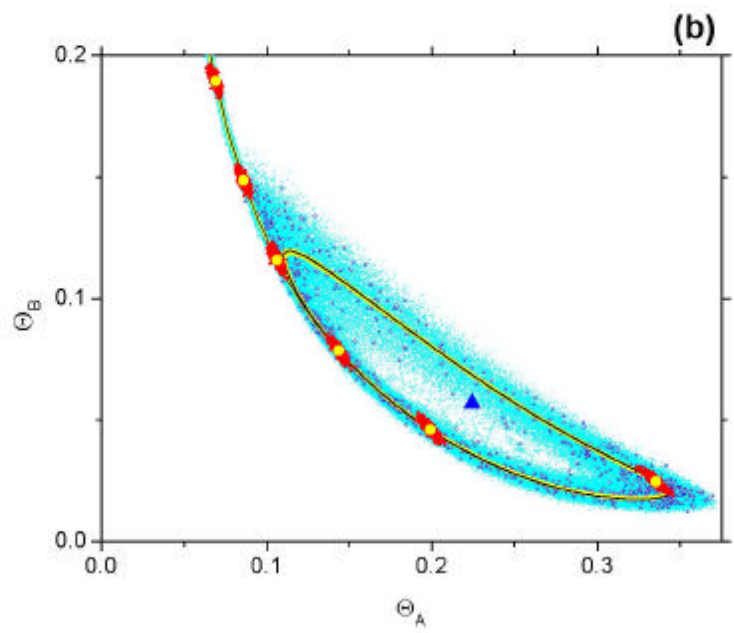